\documentclass[a4paper,12pt]{article}
\usepackage{theorem, amssymb, color}
\def\qed{{\hfill $\square$}\medskip}
\def\Tr{\mathrm{Tr}}

\newtheorem{lemma}{Lemma}

\newcommand{\<}{\langle}
\renewcommand{\>}{\rangle}

\def\proof{\noindent{\it Proof.} }

\def\bbbc{{\mathbb C}}

\def\Tr{\mathrm{Tr}\,}
\def\Diag{\mbox{Diag}\,}

\def\im{\mathrm{i}}

\def\Mn{M_n(\bbbc)}
\def\Fs0{\mathbf{F}}
\def\Fs{\mathbf{F}}
\def\As{\mathbf{A}}
\def\Bs{\mathbf{B}}
\def\Cs{\mathbf{C}}

\def\iH{{\cal H}}

\def\osum{\oplus}

\newcommand{\mathsym}[1]{{}}

\newtheorem{thm}{Theorem}
\theorembodyfont{\rm}
\newtheorem{pl}{Example}

\begin{document}
\vskip 1cm
\centerline{\LARGE {\bf Conditional SIC-POVMs}}
\bigskip\bigskip
\centerline{{\bf D\'enes Petz\footnote{E-mail: petz@math.bme.hu},
L\'aszl\'o Ruppert\footnote{E-mail: ruppertl@gmail.com} and Andr\'as Sz\'ant\'o
\footnote{E-mail: prikolics@gmail.com}}}
\bigskip
\begin{center}
Department of Mathematical Analysis, \\ Budapest University of Technology and Economics, \\ 
Egry J\'ozsef u.~1., Budapest, 1111 Hungary
\end{center}

\begin{abstract}
In this paper we examine a generalization of the symmetric informationally complete POVMs. 
SIC-POVMs are the optimal measurements for full quantum tomography, but if some parameters
of the density matrix are known, then the optimal SIC POVM should be orthogonal to a subspace.
This gives the concept of the  conditional SIC-POVM.  The existence is not known in general, 
but we give a result in the  special cases when the diagonal is known of the density matrix. 
\end{abstract}

{\bf Keywords:} Quantum state tomography, Hilbert-Schmidt distance, SIC-POVM, projections,
quasi-orthogonality.

{\bf AMS 2010 Subject Classification:} Primary 81P15; Secondary 94A15.

\section{Introduction}

The motivation for a positive operator valued measure (POVM) is in the quantum information
theory. The outcome statistics of a quantum measurement are described by (one or more) POVMs. 
A sequence of measurements on copies of a system in an unknown state will reveal the state. 
This process is called quantum state tomography \cite{Paris}.

A POVM is a set $\{E_i\,:\, 1 \le i \le k\}$ of positive operators such that 
$\sum_i E_i=I$. A quantum density matrix $\rho$ can be informed by the probability
distribution $\{\Tr \rho E_i\,:\, 1 \le i \le k\}$. A density $\rho \in \Mn$ has $n^2-1$ 
real parameters. To cover all parameters $k\ge n^2$ should hold for the POVM. We can take 
projections $P_i$, $1 \le i \le n^2$, such that
$$
\sum_{i=1}^{n^2}P_i=nI, \qquad \Tr P_iP_j=\frac{1}{n+1} \quad (i\ne j), \qquad E_i=\frac{1}{n}P_i
$$
and this is called symmetric informationally complete POVM (SIC POVM) 
by Zauner \cite{zauner} and it is rather popular now  \cite{App, App2, Col, Klapp, renes, zhu}.  
Zauner shoved the existence for $n \le 5$ and there has been  more mathemtical and numerical 
arguments \cite{Fl, scott10}. The existence of a SIC POVM is not known for every dimension. 
Another terminology for this is tight equiangular frame. We may also consider less than 
$n^2$ projections with similar properties.

A SIC POVM $\{E_i\,:\, 1 \le i \le n^2\}$ of an $n$-level system is optimal for several
arguments. For example, the SIC POVM was optimal in our paper \cite{PR} where the 
minimization of 
the determinant of the average covariance matrix was studied. Actually, this kind of 
optimization is too complicated and a different argument was in \cite{scott}, minimization of 
the square of the Hilbert-Schmidt distance of the estimation and the true density. In the
present paper the minimization of the square of the Hilbert-Schmidt distance will be used.

In this paper the subject is the state estimation again, but a part of the $n^2-1$ parameters
is supposed to be known and we want to estimate only the unknown parameters. A POVM 
$\{E_i\,:\, 1 \le i \le k\}$ is good when $k<n^2$ and $n^2-k$ parameters are known. It is 
obvious that the optimal POVM depends on the known parameters and we use the expression 
of conditional SIC POVM. This seems to be a new subject, the existence of such conditional 
SIC POVM  can be a fundamental question in different quantum tomography problems. The 
description of the conditional SIC POVM is the main result in Section 1, however, the 
existence is not at all clear. The formalism is in a finite dimensional Hilbert space $\iH$
and a state means a density matrix in $B(\iH)$. The known parameters determine a traceless
part $B \subset B(\iH)$ and the operators of the conditional SIC POVM are orthogonal to
$B$. In Section 2 a particular situation is studied, we assume that the diagonal entries of 
the state space are given. A mathematical subject called planar difference set in projective 
geometry is used there.

\section{The optimality of conditional SIC-POVMs}

We examine the case of $\Mn$. Let us suppose that $\sigma_i$ is an orthonormal basis of 
self-adjoint matrices, i.e.
$$
\sigma_i=\sigma_i^*,\quad \<\sigma_i, \sigma_j\>=\delta_{i,j}, \quad i,j \in \{0,1,2,\dots n^2-1\}.
$$
We fix $\sigma_0=\frac{1}{\sqrt{n}}I_n$. (The elements of this basis are often called  
generalized Pauli matrices.)

A quantum state $\rho$ satisfies the conditions $\Tr \rho=1$ and $\rho\ge 0$. It can
be written in the form 
$$
\rho=\sum_{i=0}^{n^2-1} \theta_i \sigma_i,
$$
where $\theta_0=\frac{1}{\sqrt{n}}$. 
A necessary condition for the coefficients can be obtained:
\begin{equation}\label{square}
\sum_{i=1}^{n^2} \theta_i^2=\Tr \rho^2 \le 1.
\end{equation}

We decompose $\Mn$ to three orthogonal subspaces:
\begin{equation}\label{decomp}
\Mn=A \osum B \osum C,
\end{equation}
where $A:=\{ \lambda I_n: \lambda \in \bbbc\}$ is one dimensional. Denote 
the orthogonal projections to the subspaces $A,B,C$ by $\As,\Bs,\Cs$.
A density matrix $\rho \in \Mn$ has the form
$$
\rho=\frac{I_n}{n}+\Bs \rho+\Cs \rho.
$$
Assume that $\Bs \rho$ is the known traceless part of $\rho$ and $\Cs \rho$ is the 
unknown traceless part of $\rho$. We use the notation $\rho_*=\rho -\Bs \rho$. The aim 
of the state estimation is to cover $\rho_*$. If the dimension of $B$ is $m$, then the 
dimension of $C$ is $n^2-m-1$. 
For the state estimation we have to use a POVM with at least $N=n^2-m$ elements. To get 
a unique solution we will use POVM with exactly $N$ elements: $\{F_1,F_2,\dots,F_N\}$. 
For obtaining optimal POVM, we will use similar arguments to \cite{qubits} which was 
a straightforward extension of the idea appeared in \cite{scott}.

If $\{Q_i\,:\, 1\le i \le  N\}$ are self-adjoint matrices satisfying the following equation
$$
\rho_*=\frac{1}{n} I+ \sum_{\sigma_i \in C} \theta_i \sigma_i=\sum_{i=1}^N  p_i Q_i, \qquad
p_i=\Tr \rho F_i,
$$
then  $\{Q_i\,:\, 1\le i \le  N\}$ is a {\bf dual frame} of $\{F_i\,:\, 1\le i \le  N\}$. Then 
the state reconstruction formula can be written as
$$
\hat{\rho_*}=\sum_{i=1}^N  \hat p_i Q_i.
$$
We define the distance as
$$
\|\rho_*-\hat{\rho_*}\|^2_2 = \Tr (\rho_*-\hat{\rho_*})^2
= \sum_{i,j=1}^N\big(p(i)-\hat{p}(i)\big)\big(p(j)-\hat{p}(j)\big)\< Q_i,Q_j\>
$$
and its expectation value is
\begin{eqnarray*}
&& \sum_{i,j=1}^N \big(p(i)\delta(i,j)-p(i)p(j)\big)\<Q_i, Q_j\> \\ && \qquad
= \sum_{i=1}^N p(i)\<Q_i, Q_i\> - \<\sum_{i=1}^N p(i)Q_i, \sum_{j=1}^N p(j) Q_j\>
\\ && \qquad =
\sum_{i=1}^N p(i)\<Q_i, Q_i\> -\Tr (\rho_*)^2.
\end{eqnarray*}
We concentrate on the first term which is
\begin{equation}\label{intdo}
\sum_{i=1}^N(\Tr F_i \rho ) \<Q_i, Q_i\>
\end{equation}
and we take the integral with respect to the Haar measure on the unitaries $\mathrm{U}(n)$.

Note first that 
$$
\int_{\mathrm{U}(n)}U P U^*\,\, d\mu (U)
$$
is the same constant $c$ for any projection of rank 1. If $\sum_{i=1}^n P_i=I_n$, then
$$
n c=\sum_{i=1}^n \int_{\mathrm{U}(n)}U P_i U^*\,\, d\mu (U)= I_n
$$
and we have $c=I_n/n$. Therefore for $A=\sum_{i=1}^n \lambda_i P_i$ we have
$$
\int_{\mathrm{U}(n)}U A U^*\,\, d\mu (U)= \sum_{i=1}^n \lambda_i c=\frac{I_n}{n}\Tr A
$$
and application to the integral of (\ref{intdo}) gives
$$
\int \Tr F_i (U\rho  U^*)\, d\mu (U)=\frac{1}{n}\Tr F_i.
$$
So we get the following quantity for the error of the state estimation:
$$
T:=\int  E\left( \|U\rho^* U^*-U\hat {\rho^*} U^*\|^2_2\right) d\mu (U)=
\frac{1}{n}\sum_{i=1}^N (\Tr F_i) \<Q_i, Q_i\>-\Tr (\rho^*)^2
$$
This is to be minimized. Since the second part is constant, our task is to minimize
the first part:
\begin{equation}\label{minprob}
\sum_{i=1}^N (\Tr F_i) \<Q_i, Q_i\> 
\end{equation}

We define the superoperator:
$$
\Fs =\sum_{i=1}^N  |F_i\>\<F_i| (\Tr F_i)^{-1}.
$$
It will have rank $N$, so if $N<n^2$ the inverse of $\Fs$ does not exists, but we can 
use its pseudo-inverse $\Fs^-$, so $\Fs^- |\sigma_i\>=0$, if $\sigma_i \in B$. $R_i$ 
is the canonical dual frame of $F_i$, if
$$
|R_i\>=\Fs^- |P_i\>,
$$
where $P_i=(\Tr F_i)^{-1}  F_i$.

\begin{lemma}
For a fixed $F_i$, (\ref{minprob}) is minimal if $Q_i=R_i$, i.e. if we use the canonical 
dual frame.
\end{lemma}

\proof
Let us use the notation $W_i=Q_i-R_i$. Then
\begin{eqnarray}
\sum_{i=1}^N \Tr F_i | R_i\>\<W_i| &= &
\sum_{i=1}^N \Tr F_i | R_i\>\<Q_i|-\sum_{i=1}^N \Tr F_i | R_i\>\<R_i| \cr &=&
\sum_{i=1}^N \Tr F_i  \Fs^- |P_i\>\<Q_i|-\sum_{i=1}^N \Tr F_i \Fs^- |P_i\>\<P_i| \Fs^- 
\cr & = &\Fs^- \sum_{i=1}^N \Tr F_i |P_i\>\<Q_i|-\Fs^- \bigg(\sum_{i=1}^N \Tr F_i  |P_i\>
\<P_i| \bigg)\Fs^- \cr &
=& \Fs^- {\bf \Pi}-\Fs^-  \Fs \Fs^- = \Fs^- {\bf \Pi} - \Fs^- {\bf \Pi}=0, \label{eR}
\end{eqnarray}
where  ${\bf \Pi}=\As+\Cs$, and we use that from
$$
|\rho^*\>=\sum_{i=1}^N  \<F_i ||\rho\>  |Q_i\>
$$
follows
$$
{\bf \Pi}=\sum_{i=1}^N  |Q_i\> \<F_i|.
$$
So we have
\begin{eqnarray*}
\sum_{i=1}^N \Tr F_i \<Q_i, Q_i\> 
&=&\sum_{i=1}^N \Tr F_i \<W_i, W_i\>+\sum_{i=1}^N \Tr F_i \<W_i, R_i\>\cr
&&
+
\sum_{i=1}^N \Tr F_i \<R_i, W_i\>+\sum_{i=1}^N \Tr F_i \<R_i, R_i\> \cr &
=&\sum_{i=1}^N \Tr F_i \<W_i, W_i\>+\sum_{i=1}^N \Tr F_i \<R_i, R_i\> 
\cr &\ge& \sum_{i=1}^N \Tr F_i \<R_i, R_i\>.
\end{eqnarray*}
\qed

We know the optimal dual frame for a fixed POVM $F_i$, and the following lemma provides a 
property for the optimal POVM:

\begin{lemma}
The quantity in (\ref{minprob}) is minimal if
$$
\Fs=\As + \frac{n-1}{N-1}\Cs.
$$
\end{lemma}

\proof
From (\ref{eR}) we have
$$
\sum_{i=1}^N (\Tr F_i)  |R(i)\>\<R(i)|=\Fs^- {\bf \Pi}=\Fs^-,
$$
so we have the equation:
$$
\sum_{i=1}^N (\Tr F_i)  \<R(i), R(i)\>=\Tr (\Fs^-).
$$
Let $\nu_1\ge \nu_2 \ge \dots \ge \nu_{n^2}$ be the eigenvalues of $\Fs$. Since the rank of 
$\Fs$ is $N$, we have $\nu_i=0$ for $i>N$. We want to minimize
$$
\Tr (\Fs^-)=\sum_{i=1}^N \frac{1}{\nu_i}.
$$
It is easy to check that $\As$ is an eigenfunction of $\Fs$ with $\nu_1=1$ eigenvalue:
$$
\Fs |I\>=\sum_{i=1}^N (\Tr F_i)  |P(i)\> \<P(i),I \>=\sum_{i=1}^N (\Tr F_i)  
|P(i)\>=\sum_{i=1}^N |F(i)\>=|I\>
$$
and we have the following condition:
$$
\sum_{i=1}^N \nu_i=\Tr \Fs =\sum_{i=1}^N \<P_i,P_i\> \Tr F_i  \le
\sum_{i=1}^N \Tr F_i=\Tr I= n.
$$
Combining these conditions we get that the measurement is optimal if $\nu_2=\nu_3=\dots
=\nu_N= \frac{n-1}{N-1}$.
\qed

Now we can obtain that the optimal POVM is a conditional SIC-POVM:

\begin{thm}\label{T:cond}
If
\begin{equation}\label{E:As}
\Fs=\As + \frac{n-1}{N-1}\Cs.
\end{equation}
then
$$
\sum_{i=1}^N P_i=\frac{N}{n}I,\qquad \Tr P_iP_j= \frac{N-n}{n(N-1)} \quad (i \ne j), \qquad
\Tr \sigma_k P_i=0 \quad(\sigma_k \in B).
$$
\end{thm}

\proof
Let us use notation $\lambda_i=\Tr F_i$, then (\ref{E:As}) has the form:
$$
\sum_{i=1}^{N}\lambda_i |P_i\>\<P_i|=\As + \frac{n-1}{N-1}\Cs.
$$

Then we have to the following equation:
\begin{equation}\label{E:As7b}
\sum_{i=1}^{N}\lambda_i \<Q|P_i\>\<P_i|Q\>=\<Q|\As +\frac{n-1}{N-1}\Cs |Q\>
\end{equation}
with  $Q := P_k - d\cdot I$.

From $\<P_i|Q\>=\Tr P_i P_k- d$ the left hand side of (\ref{E:As7b}) becomes
$$
\sum_{i=1}^{N}\lambda_i \<Q|P_i\>\<P_i|Q\>=\lambda_k (1-d)^2
+\sum_{i\ne k}\lambda_i(\Tr P_i P_k- d)^2.
$$
We can compute the right hand side as well:
$$
\As (P_k - d I)=\As P_k - d I=\As (P_k-I/n)+I/n-d I=I(1/n-d),
$$
$$
\<Q|\As |Q\>=(1/n-d) \Tr (P_k - d I)=n (1/n-d)^2 
$$
When $P_k=\sum_{i=0}^N c_i \sigma_i$, then
$$
\Cs |Q\>=\sum_{\sigma_i \in C} c_i \sigma_i,
\qquad
\<Q|\Cs |Q\>=\sum_{\sigma_i \in C} c_i^2.
$$
So (\ref{E:As7b}) becomes
\begin{equation}\label{cond1}
\lambda_k (1-d)^2
+\sum_{i\ne k}\lambda_i(\Tr P_i P_k- d)^2
=n (1/n-d)^2 +\frac{n-1}{N-1}\sum_{\sigma_i \in C} c_i^2.
\end{equation}
From (\ref{square}) we have 
\begin{equation}\label{cond2}
\sum_{\sigma_i \in C} c_i^2 \le 1-c_0^2=1-1/n.
\end{equation}
This implies
$$
\lambda_k (1-d)^2
\le n (1/n-d)^2 +\frac{n-1}{N-1}(1-1/n),
$$
which is true for every value of $d$, so
$$
\lambda_k 
\le \min_{d} \frac{n (1/n-d)^2 +\frac{n-1}{N-1}(1-1/n)}{(1-d)^2}
$$
By differentiating we can obtain that the right hand side is minimal if:
$$
d=\frac{N-n}{n (N-1)}
$$
and then we get
$$
\lambda_k \le \frac{n}{N}.
$$
Since $\sum_{i=k}^N \lambda_k=n$, we have $\lambda_1=\lambda_2=\dots
=\lambda_N=n/N$.

From that follows that there is an equality in (\ref{cond2}) too, so we have
$$
\sum_{\sigma_i \in C} c_i^2 = 1-c_0^2 \quad \Rightarrow \quad  c_i=0 \textrm{, if } 
\sigma_i \in B \quad\Rightarrow\quad 
 \Tr \sigma_i P_k=0 \textrm{, if }  \sigma_i \in B.
$$
On the other hand from (\ref{cond1}) we have
$$
\sum_{i\ne k}\frac{n}{N}\bigg(\Tr P_i P_k- \frac{N-n}{n (N-1)}\bigg)^2=0.
$$
So it implies
$$
\Tr P_i P_k=\frac{N-n}{n (N-1)}  \quad \mbox{if}\quad i\ne k. 
$$
\qed

One has to be careful about this result though, since we only consider the case 
of linear state reconstruction, as it was stated in \cite{scott}. Finding the optimal 
statistic in a more general setting requires complicated nonlinear optimalization.

Now we look at some examples related to the previous theorem and we take different $N$ values.

\begin{pl}\label{ex1}
If we do not have any information a priori about the state ($m=0,N=n^2$), then
$$
 \Tr P_iP_j= \frac{1}{n+1} \quad (i \ne j)
$$
so the optimal POVM is the well-known SIC-POVM (if it exists \cite{renes}).
\end{pl}

\begin{pl}\label{ex2}
If we know the off-diagonal elements of the state, and we want to estimate the diagonal 
entries ($m=n^2-n,N=n$), then from Theorem \ref{T:cond} it follows that the optimal POVM has 
the properties
$$
 \Tr P_iP_j= 0 \quad (i \ne j), \quad \sum_{i=1}^n P_i=I,\quad \textrm{ and }\quad  P_i 
\textrm{ is  diagonal.} 
$$
So the diagonal matrix units form an optimal POVM. \qed
\end{pl}

\begin{pl}\label{ex3}
If we know the diagonal elements of the state, and we want to estimate the off-diagonal 
entries ($m=n-1, N=n^2-n+1$), then from Theorem \ref{T:cond} it  follows that the optimal POVM 
has the properties
$$
 \Tr P_iP_j= \frac{n-1}{n^2} \quad (i \ne j), \quad 
\sum_{i=1}^n P_i=\frac{n^2-n+1}{n}I
$$
and $P_i$ has a constant diagonal. More about this case is in the next section. \qed
\end{pl}

\section{Existence of some conditional SIC-POVMs}

Theorem \ref{T:cond} tells that conditional SIC-POVMs are the optimal measurements 
if they exist, but it was not written anything about the existence of such POVMs. 
The existence of SIC-POVMs for arbitrary dimension is not known and they are a special 
case of the conditional SIC-POVMs. We can not expect to give a full description of 
SIC-POVMs, but this section contains a particular example. There are seqveral equiangular 
frames with less than $n^2$ projections \cite{etf}, but it is not clear, 
what parameters are spanned by their complementary part, ie. what the known parameters are. 
Intuition suggests that the case when the known part corresponds to a subalgebra of the 
full matrixalgebra is especially interesting.

Suppose we know the diagonal elements of a $n$-dimensional density matrix. We want to 
construct the related conditional SIC-POVM, that is subnormalized projections $P_i$ 
forming a symmetric POVM and complementary to the diagonal projections $E_i=|e_i\>\<e_i| 
\in \Mn$ ($1 \le i \le n$). These projections form a maximal abelian subalgebra.
Easy dimension counting shows, that we want to construct $N=n^2-n+1$ such projections.

So $\{|e_i\>: 1 \le i \le n\}$ is an orthonormal basis in the space. We set
\begin{equation}\label{E:q}
|\phi\>= \frac{1}{\sqrt{n}} \sum_{i=1}^{n} |e_i\>, \qquad q=e^{2\pi\im/N}
\end{equation}
and a diagonal unitary 
$$
U=\Diag(q^{\alpha_1},q^{\alpha_2},q^{\alpha_3}, \ldots q^{\alpha_n}),
$$
where the integer numbers $0 \le \alpha_i \le N-1$ are differents. Another unitary $T$ permutes 
the eigenvectors of $U$:
$$
T|e_i\>=\cases{ |e_{i+1}\>   & if  $1 \le i \le n-1$, \cr |e_1\>  &  if  $i=n$.}
$$
Note, that $T |\phi\>=T^*|\phi\>=|\phi\>$. We have
\begin{eqnarray*}
|\<U^k \phi,e_j\>|^2 &=&|\<\phi, (U^*)^k e_j\>|^2=|q^{-k\alpha_j}|^2 |\<\phi, e_j\>|^2
=|\<\phi, e_j\>|^2 \\
&=&|\<\phi, T^{j-1} e_1\>|^2=|\<(T^*)^{j-1}\phi, e_1\>|^2 =|\<\phi, e_1\>|^2
\end{eqnarray*}
and the projections $P_k:=|U^k \phi\>\<U^k \phi|$ are complementary to the diagonal projections:
$$
\Tr |U^k \phi\>\<U^k \phi|\left( |e_i\>\<e_i| -I/n\right)=0.
$$
It is easy to check that
$$
\sum_{k=1}^N \<e_i, U^k\phi\>\<U^k \phi, e_j\> = \frac{1}{n}\sum_{k=1}^N q^{-\alpha_i k}q^{\alpha_j k}
=\frac{1}{n}\sum_{k=1}^N q^{(\alpha_j-\alpha_i) k}=\frac{N}{n}\delta_{ij},
$$ 
so we obtain
$$
\sum_{k=1}^N P_k= \frac{N}{n}I
$$ 
and the sum is  multiple of $I$.

We need to choose the numbers $\alpha_1,\alpha_2,\dots, \alpha_n$ such that 
$$
\Tr P_iP_j=|\<U^i \phi|U^j \phi\>|^2=\frac{1}{n^2}\left|\sum_{m=1}^n q^{(j-i)\alpha_m}\right|^2=
\frac{1}{n^2}t
$$ 
is constant when $i \neq j$. From the formulas
$$
\sum_j \Tr P_iP_j =(N-1)\frac{1}{n^2}t +1, \qquad \sum_j \Tr P_iP_j=\Tr \left(P_i \sum_j P_j\right)=
 \frac{N}{n}
$$
we obtain $t=n-1$.

Next we use a terminology from the paper \cite{Gordon}. The set $G:=\{0,1, \dots, N-1\}$ 
is an additive group modulo $N$. The subset $D:=\{\alpha_i : 1 \le i \le n\}$ is a 
{difference set} with parameters $(N,n,\lambda)$ when the set of differences 
$\alpha_i-\alpha_j$ contains every nonzero element of $G$ exactly $\lambda$ times. When 
this holds, then we have
$$
\left|\sum_{i=1}^n q^{m\alpha_i}\right|^2=\sum_{i,j=1}^n q^{m(\alpha_i-\alpha_j)}
=n+\sum_{s=1}^{N-1} \lambda q^s=n-\lambda,
$$
where $q$ is from (\ref{E:q}). Here $\lambda=1$. If the appropriate difference set exists, 
then there exists a conditional SIC-POVM. Similar constructions of tight equiangular frames 
related to difference sets are examined in detail in \cite{diffset}.

The existence of difference sets with parameters $(N,n,1)$ is a known problem, named the 
prime power conjecture \cite{Si, Gordon}, and we get the following result:

\begin{thm}
There exists a conditional SIC-POVM with respect to the diagonal part of a density matrix 
if $n-1$ is a prime power. Then $N=n^2-n+1$ and the projection $P_i$ ($1 \le i \le N$) 
have the properties
$$
\sum_{i=1}^N P_i=\frac{N}{n}I, \qquad \Tr P_iP_j=\frac{n-1}{n^2}\quad (i\ne j).
$$
\end{thm}

A few examples about $M=\{\alpha_k: k \}$ is written here:
$$
n=2, \quad M=\{0,1\}, \qquad n=3,\quad M= \{0,1,3\},
$$ $$ n=4,\quad M=\{0,1,3,9\}, \qquad n=5, \quad M=\{0,1,4,14,16\}.
$$


\end{document}